\documentclass[12pt]{article}

\newcommand{\text}{\rm}

\begin{document}

\title{\textbf{Ghost Condensates and Dynamical Breaking of }$SL(2,R)$\textbf{\ in
Yang-Mills in the Maximal Abelian Gauge }}
\author{V.E.R. Lemes, M.S. Sarandy, and S.P. Sorella \\
{\small {\textit{UERJ - Universidade do Estado do Rio de Janeiro,}}} \\
{\small {\textit{\ Rua S\~{a}o Francisco Xavier 524, 20550-013 Maracan\~{a}, 
}}} {\small {\textit{Rio de Janeiro, Brazil.}}}}
\maketitle

\begin{abstract}
Ghost condensates of dimension two in $SU(N)$ Yang-Mills theory quantized in
the Maximal Abelian Gauge are discussed. These condensates turn out to be
related to the dynamical breaking of the $SL(2,R)$ symmetry present in this
gauge.
\end{abstract}

\vfill\newpage\ \makeatother

\renewcommand{\theequation}{\thesection.\arabic{equation}}

\section{Introduction}

Nowadays a great deal of effort is being undertaken to study condensates of
dimension two in order to improve our knowledge about the dynamics of
Yang-Mills theories in the infrared regime. For instance, the gauge
condensate $\left\langle A^{2}\right\rangle $ has been argued to be suitable
for detecting the presence of topological structures like monopoles \cite{gz}%
. An indication that the vacuum of pure Yang-Mills theory favours a
nonvanishing value of this condensate has been achieved in \cite{vkvv} by an
explicit two-loop computation of the effective potential in the Landau
gauge. A discussion of $\left\langle A^{2}\right\rangle $ in the context of
the operator product expansion and its relevance for lattice QCD may be
found in \cite{by}. Further investigations by using a recently proposed
decomposition \cite{fn} of the gauge field have been reported \cite{lf}.

An interesting mechanism providing a condensate of dimension two has also
been proposed \cite{ms, ms1,kk} in the Maximal Abelian Gauge (MAG). This
gauge, introduced by \cite{th,ksw}, has given evidences for monopoles
condensation as well as for the Abelian dominance hypothesis, which are the
key ingredients for the so called dual superconductivity \cite{conf,th}
mechanism of QCD confinement. An important point to be noted here is that
the MAG condition is nonlinear. As a consequence, a quartic ghost
interaction term must be necessarily included for renormalizability \cite
{mp,fz}. As in the case of an attractive four-fermion interaction \cite{njl}%
, this term gives rise to an effective potential resulting in a gap equation
whose nontrivial solution at weak coupling yields a nonvanishing
off-diagonal ghost-antighost condensate $\left\langle \overline{c}%
c\right\rangle $ of dimension two. The physical relevance of this ghost
condensate lies in the fact that it is believed to be part of a more general
two-dimensional condensate, namely $\left( \frac{1}{2}\left\langle A_{\mu
}^{a}A_{\mu }^{a}\right\rangle -\xi \left\langle \overline{c}%
^{a}c^{a}\right\rangle \right) $, where $\xi $ denotes the gauge parameter
of the MAG and the index $a$ runs over all the off-diagonal generators. This
condensate has been introduced due to its BRST\ invariance \cite{ope} and it
is expected to provide effective masses for both off-diagonal gauge and
ghost fields \cite{ope, dd,dd1}.

Besides the computation of the effective potential, the problem of
identifying the symmetry which is dynamically broken by the ghost
condensation has also begun to be faced \cite{ms, ms1,kk}. In the case of $%
SU(2),$ the ghost condensation has been interpreted as a breaking of a
global $SL(2,R)$ symmetry \cite{ms,ms1} displayed by Yang-Mills in the MAG.
In ref.\cite{kk} the one-loop effective potential for the ghost condensation
in the case of $SU(3)$ has been computed. Recently, the authors \cite{sl2r}
have been able to establish the existence of $SL(2,R)$ in the MAG, for the
general case of $SU(N)$.

The aim of the present work is to continue the investigation on the ghost
condensation and their relationship with the dynamical symmetry breaking of $%
SL(2,R)$, for the general case of $SU(N)$. As already observed \cite{sl2r},
the breaking of $SL(2,R)$ can actually occur in different channels,
according to which generators are broken. More specifically, the three
generators of $SL(2,R)$, namely $\delta $, $\overline{\delta }$ and $\delta
_{FP}\;$are known \cite{oj} to obey the algebra $\left[ \delta ,\overline{%
\delta }\right] =$ $\delta _{FP}$, where $\delta _{FP}$ denotes the ghost
number.

The condensate $\left\langle \overline{c}c\right\rangle $ analysed in \cite
{ms, ms1,kk} corresponds to the breaking of the generators $\delta $, $%
\overline{\delta }$. In this paper we shall analyse the other off-diagonal
condensates $\left\langle c\;c\right\rangle $ and $\left\langle \overline{c}%
\;\overline{c}\right\rangle $ which are related to the breaking of $\left(
\delta ,\delta _{FP}\right) $ and of $\left( \overline{\delta },\delta
_{FP}\right) $, respectively \cite{sl2r}. We remark also that the existence
of different channels for the ghost condensation has an analogy in
superconductivity, known as the BCS\footnote{%
Particle-particle and hole-hole pairing.} versus the Overhauser\footnote{%
Particle-hole pairing.} effect \cite{ov}. In the present case the
Faddeev-Popov charged condensates $\left\langle c\;c\right\rangle $ and $%
\left\langle \overline{c}\;\overline{c}\right\rangle $ would correspond to
the BCS channel, while $\left\langle \overline{c}c\right\rangle $ to the
Overhauser channel. Therefore, the possibility of describing the ghost
condensation in the case of $SU(N)$ as a dynamical symmetry breaking of the
ghost number seems to be rather natural.

The paper is organized as follows. In Section 2 a brief review of the
quantization of $SU(N)$ Yang-Mills in the MAG is provided. In Section 3 the
dynamical symmetry breaking of the ghost number in the case of $SU(2)$ is
discussed in detail. Section 4 is devoted to the generalization to $SU(N),$
analysing, in particular, the case of $SU(3)$. In the last Section the
conclusions are presented.

\section{Yang-Mills theory in the MAG}

Let $\mathcal{A}_{\mu }$ be the Lie algebra valued connection for the gauge
group $SU(N),$ whose generators $T^{A}\,,\;\left[ T^{A},T^{B}\right]
=f^{ABC}T^{C}\,$, are chosen to be antihermitean and to obey the
orthonormality condition $\mathrm{Tr}\left( T^{A}T^{B}\right) =\delta ^{AB}$%
, with $A,B,C=1,..,\left( N^{2}-1\right) $. Following \cite{th,ksw}, we
decompose the gauge field into its off-diagonal and diagonal parts, namely 
\begin{equation}
\mathcal{A}_{\mu }=\mathcal{A}_{\mu }^{A}T^{A}=A_{\mu }^{a}T^{a}+A_{\mu
}^{i}T^{\,i},  \label{conn-sun}
\end{equation}
where the index $i$ labels the $N-1$ generators $T^{\,i}$ of the Cartan
subalgebra. The remaining $N(N-1)$ off-diagonal generators $T^{a}$ will be
labelled by the index $a$. Accordingly, the field strength decomposes as 
\begin{equation}
\mathcal{F}_{\mu \nu }=\mathcal{F}_{\mu \nu }^{A}T^{A}=F_{\mu \nu
}^{a}T^{a}+F_{\mu \nu }^{i}T^{\,i}\;,  \label{fsn}
\end{equation}
with the off-diagonal and diagonal parts given respectively by 
\begin{eqnarray}
F_{\mu \nu }^{a} &=&D_{\mu }^{ab}A_{\nu }^{b}-D_{\nu }^{ab}A_{\mu
}^{b}\;+g\,f^{abc}A_{\mu }^{b}A_{\nu }^{c}\;,  \nonumber \\
F_{\mu \nu }^{i} &=&\partial _{\mu }A_{\nu }^{i}-\partial _{\nu }A_{\mu
}^{i}+gf^{abi}A_{\mu }^{a}A_{\nu }^{b}\;\,,  \label{fscompn}
\end{eqnarray}
where the covariant derivative $D_{\mu }^{ab}$ is defined with respect to
the diagonal components $A_{\mu }^{i}$ 
\begin{equation}
D_{\mu }^{ab}\equiv \partial _{\mu }\delta ^{ab}-gf^{abi}A_{\mu
}^{i}\,\,\,\,\,\,.  \label{cdern}
\end{equation}
For the Yang-Mills action one obtains 
\begin{equation}
S_{\mathrm{YM}}=-\frac{1}{4}\int d^{4}x\,\left( F_{\mu \nu }^{a}F^{a\mu \nu
}+F_{\mu \nu }^{i}F^{i\mu \nu }\right) \;.  \label{symn}
\end{equation}
The so called MAG gauge condition \cite{th,ksw} amounts to fix the value of
the covariant derivative $(D_{\mu }^{ab}A^{b\mu })$ of the off-diagonal
components. However, this condition being nonlinear, a quartic ghost
self-interaction term is required. Following \cite{ms,ms1,kk}, the
corresponding gauge fixing term turns out to be 
\begin{equation}
S_{\mathrm{MAG}}=s\,\int d^{4}x\,\left( \overline{c}^{a}\left( D_{\mu
}^{ab}A^{b\mu }+\frac{\xi }{2}b^{a}\right) -\frac{\xi }{2}gf\,^{abi}%
\overline{c}^{a}\overline{c}^{b}c^{i}-\frac{\xi }{4}gf\,^{abc}c^{a}\overline{%
c}^{b}\overline{c}^{c}\right) \;,  \label{smn}
\end{equation}
where  $\xi $ is the gauge parameter and $s$ denotes the nilpotent BRST
operator acting as 
\begin{eqnarray}
sA_{\mu }^{a} &=&-\left( D_{\mu }^{ab}c^{b}+gf^{\,abc}A_{\mu
}^{b}c^{c}+gf^{\,abi}A_{\mu }^{b}c^{i}\right) ,\,\,\,\;sA_{\mu }^{i}=-\left(
\partial _{\mu }c^{i}+gf\,^{iab}A_{\mu }^{a}c^{b}\right) \;,  \nonumber \\
sc^{a} &=&gf\,^{abi}c^{b}c^{i}+\frac{g}{2}f\,^{abc}c^{b}c^{c},\,\,\,\,\,\,\,%
\,\,\,\,\,\,\,\,\,\,\,\,\,\,\,\,\,\,\,\,\,\,\,\,\,\,\,\,\,\,\,\,\,sc^{i}=%
\frac{g}{2}\,f\,^{iab}c^{a}c^{b},  \nonumber \\
s\overline{c}^{a}
&=&b^{a}\;,\,\,\,\,\,\,\,\,\,\,\,\,\,\,\,\,\,\,\,\,\,\,\,\,\,\,\,\,\,\,\,\,%
\,\,\,\,\,\,\,\,\,\,\,\,\,\,\,\,\,\,\,\,\,\,\,\,\,\,\,\,\,\,\,\,\,\,\,\,\,\,%
\,\,\,\,\,\,\,\,\,\,\,\,\,\,\,\,s\overline{c}^{i}=b^{i}\;,  \nonumber \\
sb^{a}
&=&0\;,\,\,\,\,\,\,\,\,\,\,\,\,\,\,\,\,\,\,\,\,\,\,\,\,\,\,\,\,\,\,\,\,\,\,%
\,\,\,\,\,\,\,\,\,\,\,\,\,\,\,\,\,\,\,\,\,\,\,\,\,\,\,\,\,\,\,\,\,\,\,\,\,\,%
\,\,\,\,\,\,\,\,\,\,\,\,\,\,\,\,\,\,sb^{i}=0\;.  \label{BRSTn}
\end{eqnarray}
Here $c^{a},c^{i}$ are the off-diagonal and the diagonal components of the
Faddeev-Popov ghost field, while $\overline{c}^{a},b^{a}$ are the
off-diagonal antighost and Lagrange multiplier. We also observe that the
BRST\ transformations $\left( \ref{BRSTn}\right) $ have been obtained by
their standard form upon projection on the off-diagonal and diagonal
components of the fields. Concerning the gauge parameters, we remark that,
in general, the MAG condition allows for the introduction of two independent
parameters \cite{mp}, while in eq.$\left( \ref{smn}\right) $ a unique gauge
parameter $\xi $ has been introduced. However, the resulting theory turns
out to be renormalizable due to the existence of a further Ward identity
which ensures the stability under radiative corrections \cite{fz}.
Expression $\left( \ref{smn}\right) $ is easily worked out and yields 
\begin{eqnarray}
S_{\mathrm{MAG}} &=&\int d^{4}x\left( b^{a}\left( D_{\mu }^{ab}A^{b\mu }+%
\frac{\xi }{2}b^{a}\right) +\overline{c}^{a}D_{\mu }^{ab}D^{\mu bc}c^{c}+g%
\overline{c}^{a}f^{abi}\left( D_{\mu }^{bc}A^{c\mu }\right) c^{i}\right.  
\nonumber \\
&&\,\,\,\,\,\,\,\,\,\,\,\,\,\,\,\,\,\,+g\overline{c}^{a}D_{\mu }^{ab}\left(
f^{bcd}A^{c\mu }c^{d}\right) -g^{2}f^{abi}f^{cdi}\overline{c}^{a}c^{d}A_{\mu
}^{b}A^{c\mu }-\xi gf^{abi}b^{a}\overline{c}^{b}c^{i}\;  \nonumber \\
&&\,\,\,\,\,\,\,\,\,\,\,\,\,\,\,\,\,-\frac{\xi }{2}gf^{abc}b^{a}\overline{c}%
^{b}c^{c}-\frac{\xi }{4}g^{2}f^{abi}f^{cdi}\overline{c}^{a}\overline{c}%
^{b}c^{c}c^{d}-\frac{\xi }{4}g^{2}f^{abc}f^{adi}\overline{c}^{b}\overline{c}%
^{c}c^{d}c^{i}  \nonumber \\
&&\left. \,\,\,\,\,\,\,\,\,\,\,\,\,\,-\frac{\xi }{8}g^{2}f^{abc}f^{ade}%
\overline{c}^{b}\overline{c}^{c}c^{d}c^{e}\right) \;.  \label{smn2}
\end{eqnarray}
Notice also that for positive values of $\xi $ the quartic ghost interaction
is attractive. As it has been shown in \cite{ms,ms1,kk}, the formation of
ghost condensates at weak coupling is thus favoured.

The MAG condition allows for a residual local $U(1)^{N-1}$ invariance with
respect to the diagonal subgroup, which has to be fixed by means of a
suitable further gauge condition on the diagonal components $A_{\mu }^{i}$
of the gauge field. Adopting a covariant Landau condition, the remaining
gauge fixing term is given by 
\begin{equation}
S_{\mathrm{diag}}=s\,\int d^{4}x\,\;\overline{c}^{i}\partial _{\mu }A^{i\mu
}\;=\int d^{4}x\,\;\left( b^{i}\partial _{\mu }A^{i\mu }+\overline{c}%
^{i}\partial ^{\mu }\left( \partial _{\mu }c^{i}+gf\,^{iab}A_{\mu
}^{a}c^{b}\right) \right) \;,  \label{lg}
\end{equation}
where $\overline{c}^{i},b^{i}$ are the diagonal antighost and Lagrange
multiplier. As in the familiar case of $QED$, the diagonal gauge fixing
gives rise to a linearly broken $U(1)^{N-1}$ Ward identity \cite{fz} which
takes the form 
\begin{equation}
\mathcal{W}^{i}S=-\partial ^{2}b^{i}\;,\;\;\;\;\;\;\mathcal{W}^{i}=\partial
_{\mu }\frac{\delta }{\delta A_{\mu }^{i}}+gf^{abi}\left( A_{\mu }^{a}\frac{%
\delta }{\delta A_{\mu }^{b}}+c^{a}\frac{\delta }{\delta c^{b}}+b^{a}\frac{%
\delta }{\delta b^{b}}+\overline{c}^{a}\frac{\delta }{\delta \overline{c}^{b}%
}\right) \;,  \label{gh-u1}
\end{equation}
where $S=S_{\mathrm{YM}}+S_{\mathrm{MAG}}+S_{\mathrm{diag}}$. From $\left( 
\ref{gh-u1}\right) $ one sees that the diagonal components $A_{\mu }^{i}$ of
the gauge field play the role of massless photons, while all off-diagonal
components behave as charged matter fields.

\section{ Dynamical ghost number symmetry breaking: the case of SU(2)}

In this Section we discuss the dynamical mechanism which, due to the quartic
ghost interaction term, leads to the existence of the off-diagonal
condensates $\left\langle c\mathrm{\ }c\right\rangle $ and $\left\langle 
\overline{c}\mathrm{\ }\overline{c}\right\rangle .$ These condensates will
realize a dynamical breaking of the ghost number symmetry. In the case of $%
SU(2)$ the gauge fixing term $\left( \ref{smn2}\right) $ simplifies to 
\begin{eqnarray}
S_{\mathrm{MAG}} &=&\int d^{4}x\left( b^{a}\left( D_{\mu }^{ab}A^{b\mu }+%
\frac{\xi }{2}b^{a}\right) +\overline{c}^{a}D_{\mu }^{ab}D^{\mu bc}c^{c}+g%
\overline{c}^{a}\varepsilon ^{ab}\left( D_{\mu }^{bc}A^{c\mu }\right)
c\right.   \nonumber \\
&&\left. \,\,\,\,\,\,\,\,\,\,\,\,\,\,\,\,\,\,-g^{2}\varepsilon
^{ab}\varepsilon ^{cd}\overline{c}^{a}c^{d}A_{\mu }^{b}A^{c\mu }-\xi
g\varepsilon ^{ab}b^{a}\overline{c}^{b}c-\frac{\xi }{4}g^{2}\varepsilon
^{ab}\varepsilon ^{cd}\overline{c}^{a}\overline{c}^{b}c^{c}c^{d}\right) \; 
\nonumber \\
&&  \label{su2}
\end{eqnarray}
where $\varepsilon ^{ab}=\varepsilon ^{ab3}\;(a,b=1,2)$ are the off-diagonal
components of the $SU(2)$ structure constants $\varepsilon ^{ABC}$, $c=c^{3}$
is the diagonal ghost field, and $D_{\mu }^{ab}=\left( \partial _{\mu
}\delta ^{ab}-g\varepsilon ^{ab}A_{\mu }\right) \,\,\,\,$is the covariant
derivative, with $A_{\mu }=A_{\mu }^{3}$ denoting the diagonal component of
the gauge connection. In order to deal with the quartic ghost interaction we
linearize it by introducing a pair of real$\footnote{%
This property follows from the hermiticity properties of the ghost and
antighost fields, chosen here as in \cite{kon}.}$ auxiliary
Hubbard-Stratonovich fields $\left( \varphi ,\overline{\varphi }\right) $,
so that 
\begin{equation}
-\frac{\xi }{4}g^{2}\varepsilon ^{ab}\varepsilon ^{cd}\overline{c}^{a}%
\overline{c}^{b}c^{c}c^{d}\longrightarrow -\frac{1}{\xi g^{2}}\overline{%
\varphi }\varphi +\frac{1}{2}\varphi \varepsilon ^{ab}\overline{c}^{a}%
\overline{c}^{b}-\frac{1}{2}\overline{\varphi }\varepsilon ^{ab}c^{a}c^{b}\;.
\label{hssu2}
\end{equation}
The invariance of the gauge fixed action $S$ under the BRST\ transformation
is guaranteed by demanding that 
\begin{equation}
s\overline{\varphi }=\xi g^{2}\varepsilon ^{ab}b^{a}\overline{c}%
^{b\;},\;\;\;\;\;\;\;s\varphi =0\;.  \label{sf}
\end{equation}
>From the expression $\left( \ref{hssu2}\right) $ one sees that the
requirement of positivity of the gauge fixing parameter, \textit{i.e. }$\xi
>0$, will ensure that the effective potential $V_{eff}\,(\varphi ,\overline{%
\varphi })$ for the Hubbard-Stratonovich fields $\left( \varphi ,\overline{%
\varphi }\right) $ will be bounded from below, a necessary physical
requirement. Moreover, in the following we shall see that a nontrivial
vacuum configuration, corresponding to a nonvanishing ghost condensation,
will be obtained by setting $\xi =22/3$.

According to our present aim, the auxiliary fields $\left( \varphi ,%
\overline{\varphi }\right) $ carry a nonvanishing Faddeev-Popov charge, as
it can be seen from the following table where the dimension and the ghost
number of all fields are displayed. 
\begin{eqnarray*}
&& 
\begin{tabular}{|l|l|l|l|l|l|l|l|l|}
\hline
$\mathrm{Field}$ & $A_{\mu }^{a}$ & $A_{\mu }$ & $c^{a}$ & $c$ & $\overline{c%
}^{a}$ & $b^{a}$ & $\varphi $ & $\overline{\varphi }$ \\ \hline
$\mathrm{Gh.Number}$ & 0 & 0 & 1 & 1 & -1 & 0 & 2 & -2 \\ \hline
$\mathrm{Dimension}$ & 1 & 1 & 1 & 1 & 1 & 2 & 2 & 2 \\ \hline
\end{tabular}
\\
&&_{\mathrm{Table\,1.\,Ghost\,number\,\,and\,\,canonical\,\,dimension\,\,of%
\,\,the\,\,fields\,}}
\end{eqnarray*}
Therefore, a nonvanishing vacuum expectation value for $\left( \varphi ,%
\overline{\varphi }\right) $ will have the meaning of a breaking of the
ghost number generator of $SL(2,R)$. In order to analyse whether a
nontrivial vacuum for $\left( \varphi ,\overline{\varphi }\right) $ is
selected, we follow the Coleman-Weinberg procedure \cite{cw} and evaluate
the one-loop effective potential $V_{eff}\,(\varphi ,\overline{\varphi })$
for constant configurations of $\left( \varphi ,\overline{\varphi }\right) $%
. A straightforward computation gives 
\begin{equation}
V_{eff}\,(\varphi ,\overline{\varphi })=\frac{\overline{\varphi }\varphi }{%
\xi g^{2}}+i\int \frac{d^{4}k}{(2\pi )^{4}}\ln \left( (-k^{2})^{2}+\overline{%
\varphi }\varphi \right) \;.  \label{veff}
\end{equation}
Using the dimensional regularization and adopting the renormalization
condition at arbitrary scale $M$ 
\begin{equation}
\left. {\frac{{\partial ^{2}V_{eff}}}{{\partial \overline{\varphi }\partial
\varphi }}}\right| _{\overline{\varphi }\varphi =M^{4}}=\frac{1}{\xi g^{2}}%
\;,  \label{nc}
\end{equation}
the renormalized effective potential is found to be 
\begin{equation}
V_{eff}\,(\varphi ,\overline{\varphi })=\overline{\varphi }\varphi \left( 
\frac{1}{\xi g^{2}}+\frac{1}{32\pi ^{2}}\left( \ln \frac{\overline{\varphi }%
\varphi }{M^{4}}-2\right) \right) \;.  \label{rveff2}
\end{equation}
The minimization of $V_{eff}$ yields the condition 
\begin{equation}
\ln \frac{\overline{\varphi }\varphi }{M^{4}}=1-\frac{32\pi ^{2}}{\xi g^{2}}%
\;,  \label{mveff2}
\end{equation}
which gives the nontrivial vacuum configuration 
\begin{equation}
\overline{\varphi }\equiv \overline{v}=\overline{\beta }\,M^{2}\,\exp \left( 
\frac{1}{2}-\frac{16\pi ^{2}}{\xi g^{2}}\right) \,\,,\,\;\;\;\,\varphi
\equiv v=\beta \,M^{2}\,\exp \left( \frac{1}{2}-\frac{16\pi ^{2}}{\xi g^{2}}%
\right) \;,  \label{ntv}
\end{equation}
where $\beta $ and $\overline{\beta }$ are dimensionless constants with
ghost number $(2,-2)$, obeying the constraint $\beta \overline{\beta }=1.\;$%
Of course, their introduction accounts for $\left( \varphi ,\overline{%
\varphi }\right) $ being Faddeev-Popov charged. From eq.$\left( \ref{hssu2}%
\right) $ one sees that the nonvanishing expectation value of $\left(
\varphi ,\overline{\varphi }\right) $ leads to the existence of the ghost
condensates $\left\langle \varepsilon ^{ab}c^{a}c^{b}\right\rangle \;$and $%
\left\langle \varepsilon ^{ab}\overline{c}^{a}\overline{c}^{b}\right\rangle $%
.

$\;$In order to analyse the consequences following from the nontrivial
ground state configuration $\left( \ref{ntv}\right) $ let us look at the
ghost propagators in the condensed vacuum. They are easily computed and read 
\begin{eqnarray}
\left\langle c^{a}(p)\;c^{b}(-p)\right\rangle &=&i\frac{v\varepsilon ^{ab}}{%
(p^{2})^{2}+\overline{v}v},\;\;\;\;\;\;\;\;\;\left\langle \overline{c}%
^{a}(p)\;\overline{c}^{b}(-p)\right\rangle =-i\frac{\overline{v}\varepsilon
^{ab}}{(p^{2})^{2}+\overline{v}v}\;.  \nonumber \\
\left\langle \overline{\,c}^{a}(p)\;c^{b}(-p)\right\rangle &=&i\frac{%
p^{2}\,\,\delta ^{ab}}{(p^{2})^{2}+\overline{v}v}\,\,\,\,\,,  \label{prop}
\end{eqnarray}
One sees thus that, due to the existence of the condensates $\left\langle
\varepsilon ^{ab}c^{a}c^{b}\right\rangle \;$and $\left\langle \varepsilon
^{ab}\overline{c}^{a}\overline{c}^{b}\right\rangle $, the propagators $%
\left( \ref{prop}\right) $ become regular in the low energy region, the
infrared cutoff being given by $\overline{v}v$. Moreover, from expressions $%
\left( \ref{prop}\right) $, it follows that another condensate $\left\langle 
\overline{\,c}^{a}c^{a}\right\rangle $ of ghost number zero is nonvanishing,
namely 
\begin{equation}
\left\langle \overline{\,c}^{a}c^{a}\right\rangle =-\frac{\left( \overline{v}%
v\right) ^{1/2}}{16\pi }\;.  \label{cbc}
\end{equation}
We remark the absence in the propagator $\left\langle \overline{\,c}%
^{a}(p)\;c^{b}(-p)\right\rangle $ of a term containing the antisymmetric
tensor $\varepsilon ^{ab}$, forbidding the presence of the condensate $%
\left\langle \varepsilon ^{ab}c^{a}\overline{c}^{b}\right\rangle $. Notice
also that all ghost condensates $\left\langle \varepsilon
^{ab}c^{a}c^{b}\right\rangle $ , $\left\langle \varepsilon ^{ab}\overline{c}%
^{a}\overline{c}^{b}\right\rangle ,$ and $\left\langle \overline{\,c}%
^{a}c^{a}\right\rangle $ are invariant under the residual $U(1)$
transformations, meaning that the corresponding Ward identity $\left( \ref
{gh-u1}\right) $ remains unbroken.

Let us now turn to analyse the stability within the perturbative framework
of the vacuum solution $\left( \ref{ntv}\right) $. Consistency with the
one-loop computation requires that the vacuum configuration $\left( \ref{ntv}%
\right) $ is a solution of the gap equation $\left( \ref{mveff2}\right) $ at
arbitrary small coupling $g$, ensuring that the logarithmic contributions
for the effective potential are small and therefore compatible with the
perturbative expansion \cite{cw}. This condition will fix the order of
magnitude of $\left( \overline{v}v\right) $. In order to solve the equation $%
\left( \ref{mveff2}\right) $ at small coupling we introduce the
renormalization group invariant QCD scale parameter $\Lambda _{\mathrm{QCD}}$
\begin{equation}
\Lambda _{\mathrm{QCD}}^{2}=M^{2}\exp \left( -\frac{16\pi ^{2}}{\beta
_{0}g^{2}}\right) \;,  \label{lqcd}
\end{equation}
where $\beta _{0}$ is the 1-loop coefficient of the $\beta $-function of
pure Yang-Mills 
\begin{equation}
\beta (g)=-\beta _{0}\frac{g^{3}}{16\pi ^{2}}+O(g^{5})\,\,\,\,,\,\,\,\;\;\;%
\;\beta _{0}=\frac{11}{3}\,N\,,\,\;\;\mathrm{{for\;\;\;}SU(N)\;\;.}
\label{betaf}
\end{equation}
Inserting $\left( \ref{lqcd}\right) $ in the gap equation $\left( \ref
{mveff2}\right) $ one gets 
\begin{equation}
\ln \frac{\overline{\varphi }\varphi }{\Lambda _{\mathrm{QCD}}^{4}}=\frac{%
32\pi ^{2}}{g^{2}}\left( \frac{1}{\beta _{0}}-\frac{1}{\xi }\right) +1\;.
\label{lqcd2}
\end{equation}
Therefore, according to \cite{ms, ms1,kk}, the existence of a solution at
arbitrary small coupling is ensured by choosing for the gauge parameter $\xi 
$ the value 
\begin{equation}
\xi =\beta _{0}=\frac{22}{3}\;.  \label{v}
\end{equation}
It is worth mentioning here that, as shown in \cite{ms, ms1,kk}, the
one-loop anomalous dimensions $\gamma _{\varphi }$, $\gamma _{\overline{%
\varphi }}$ of the auxiliary fields $\overline{\varphi }$, $\varphi $ turn
out to vanish when the value of $\xi $ is precisely that of eq.$\left( \ref
{v}\right) $. In turn, this ensures that the one-loop effective potential $%
\left( \ref{rveff2}\right) $ obeys the renormalization group equations.
Also, as a consequence of eq.$\left( \ref{v}\right) $, the breaking $\left( 
\overline{v}v\right) ^{1/2}$ turns out to be of the order of $\Lambda _{%
\mathrm{QCD}}^{2}$. Concerning the symmetry breaking aspects related to the
existence of the ghost condensates $\left\langle \varepsilon
^{ab}c^{a}c^{b}\right\rangle $, $\left\langle \varepsilon ^{ab}\overline{c}%
^{a}\overline{c}^{b}\right\rangle $, it is apparent that a nonvanishing
expectation value for the Faddeev-Popov charged auxiliary fields $\varphi $
and $\overline{\varphi }$ leads to a breaking of the ghost number. Let us
proceed now with the generalization to the case of $SU(N).$

\section{Generalization to SU(N)}

In order to generalize the previous mechanism to $SU(N)$ we introduce a set
of real Faddeev-Popov charged auxiliary fields $\varphi ,\overline{\varphi }$
in the adjoint representation, namely 
\begin{equation}
\overline{\varphi }=\overline{\varphi }^{A}T^{A}=\overline{\varphi }%
^{a}T^{a}+\overline{\varphi }^{i}T^{i}\;,\;\;\;\;\;\;\;\;\;\;\varphi
=\varphi ^{A}T^{A}=\varphi ^{a}T^{a}+\varphi ^{i}T^{i}\;,  \label{pbpn}
\end{equation}
where the indices $a$ and $i$ run over the off-diagonal and diagonal
generators, respectively. The pure off-diagonal ghost terms of the gauge
fixing $\left( \ref{smn2}\right) $ can be rewritten as 
\begin{eqnarray}
S_{\mathrm{MAG}}^{\mathrm{off}} &=&\int d^{4}x\left( \overline{c}%
^{a}\partial ^{2}c^{a}-\frac{1}{\xi g^{2}}\varphi ^{i}\overline{\varphi }%
^{i}+\frac{1}{2}\varphi ^{i}f^{abi}\overline{c}^{a}\overline{c}^{b}-\frac{1}{%
2}\overline{\varphi }^{i}f^{abi}c^{a}c^{b}\right.  \nonumber \\
&&\,\,\,\,\,\,\,\,\,\,\,\,\,\,\,\,\,\,\left. -\frac{1}{\xi g^{2}}\overline{%
\varphi }^{a}\varphi ^{a}+\frac{1}{2\sqrt{2}}\varphi ^{a}f^{abc}\overline{c}%
^{b}\overline{c}^{c}-\frac{1}{2\sqrt{2}}\overline{\varphi }%
^{a}f^{abc}c^{b}c^{c}\right) \;  \label{hsn}
\end{eqnarray}
With the introduction of the auxiliary fields $\varphi ^{i}$, $\overline{%
\varphi }^{i}$, $\overline{\varphi }^{a}$, $\varphi ^{a}$ the Ward identity $%
\left( \ref{gh-u1}\right) $ generalizes to 
\begin{equation}
\mathcal{W}^{i}S=-\partial ^{2}b^{i}\;,  \label{sunw}
\end{equation}
where 
\begin{equation}
\mathcal{W}^{i}=\partial _{\mu }\frac{\delta }{\delta A_{\mu }^{i}}%
+gf^{abi}\left( A_{\mu }^{a}\frac{\delta }{\delta A_{\mu }^{b}}+c^{a}\frac{%
\delta }{\delta c^{b}}+b^{a}\frac{\delta }{\delta b^{b}}+\overline{c}^{a}%
\frac{\delta }{\delta \overline{c}^{b}}+\varphi ^{a}\frac{\delta }{\delta
\varphi ^{b}}+\overline{\varphi }^{a}\frac{\delta }{\delta \overline{\varphi 
}^{b}}\right) \;  \label{sunwop}
\end{equation}
According to this identity, only the $U(1)^{N-1}-$invariant diagonal fields $%
\varphi ^{i},\overline{\varphi }^{i}$ may acquire a nonvanishing vacuum
expectation value. As before, let us look at the one-loop effective
potential, which in the present case reads 
\begin{equation}
V_{eff}\,(\varphi ^{i},\overline{\varphi }^{i})=\frac{\overline{\varphi }%
^{i}\varphi ^{i}}{\xi g^{2}}+\frac{i}{2}\;\ln \det \mathcal{M}^{ab}\;,
\label{veffsun}
\end{equation}
where $\mathcal{M}^{ab}$ denotes the $\left( 2N(N-1)\right) \times \left(
2N(N-1)\right) $ matrix 
\begin{equation}
\mathcal{M}^{ab}=\left( 
\begin{array}{ll}
f^{abi}\varphi ^{i} & \delta ^{ab}\partial ^{2} \\ 
-\delta ^{ab}\partial ^{2} & -f^{abi}\overline{\varphi }^{i}
\end{array}
\right) \;.  \label{mx}
\end{equation}
In the case of $SU(3)$, the Cartan subgroup has dimension two, with $\varphi
^{i}=\left( \varphi ^{3},\varphi ^{8}\right) $ and $\overline{\varphi }%
^{i}=\left( \overline{\varphi }^{3},\overline{\varphi }^{8}\right) .$ Making
use of the explicit values of the structure constants, the effective
potential $\left( \ref{veffsun}\right) $ is found to be 
\begin{equation}
V_{eff}\,(\varphi ^{i},\overline{\varphi }^{i})=\frac{\overline{\varphi }%
^{i}\varphi ^{i}}{\xi g^{2}}+i\sum_{\alpha =1}^{3}\int \frac{d^{4}k}{(2\pi
)^{4}}\ln \left( (-k^{2})^{2}+\left( \varepsilon _{\alpha }^{i}\overline{%
\varphi }^{i}\right) \left( \varepsilon _{\alpha }^{j}\varphi ^{j}\right)
\right) \;,  \label{veffsu3}
\end{equation}
where $\varepsilon _{\alpha }$ are the root vectors of $SU(3)$, given by $%
\varepsilon _{1}=(1,0)$, $\varepsilon _{2}=(-1/2,-\sqrt{3}/2)$ and $%
\varepsilon _{3}=(-1/2,\sqrt{3}/2)$. We observe that expression $\left( \ref
{veffsun}\right) $, although obtained in a different way, is very similar to
that of \cite{kk}. The effective potential $\left( \ref{veffsun}\right) $
turns out to possess global minima along the directions of the roots, given
by the following configurations $\left( \;\varphi _{\alpha }^{3},\overline{%
\varphi }_{\alpha }^{3},\varphi _{\alpha }^{8},\overline{\varphi }_{\alpha
}^{8}\right) \;$ 
\begin{eqnarray}
\varphi _{1}^{3} &=&2^{1/3}\beta M^{2}\exp \left( \frac{1}{2}-\frac{32\pi
^{2}}{3\xi g^{2}}\right) ,\;\overline{\varphi }_{1}^{3}=2^{1/3}\overline{%
\beta }M^{2}\exp \left( \frac{1}{2}-\frac{32\pi ^{2}}{3\xi g^{2}}\right)
\;,\;\;\;\;\;  \nonumber \\
\varphi _{1}^{8} &=&\overline{\varphi }_{1}^{8}=0\;,\;  \label{r1}
\end{eqnarray}

\begin{eqnarray}
\varphi _{2}^{3} &=&4^{-1/3}\beta M^{2}\exp \left( \frac{1}{2}-\frac{32\pi
^{2}}{3\xi g^{2}}\right) ,\;\overline{\varphi }_{2}^{3}=4^{-1/3}\overline{%
\beta }M^{2}\exp \left( \frac{1}{2}-\frac{32\pi ^{2}}{3\xi g^{2}}\right) \;,
\nonumber \\
\varphi _{2}^{8} &=&\sqrt{3}\varphi _{2}^{3},\,\,\overline{\varphi }_{2}^{8}=%
\sqrt{3}\;\overline{\varphi }_{2}^{3}\;,\;  \label{r2}
\end{eqnarray}
and 
\begin{eqnarray}
\varphi _{3}^{3} &=&4^{-1/3}\beta M^{2}\exp \left( \frac{1}{2}-\frac{32\pi
^{2}}{3\xi g^{2}}\right) ,\;\overline{\varphi }_{3}^{3}=4^{-1/3}\overline{%
\beta }M^{2}\exp \left( \frac{1}{2}-\frac{32\pi ^{2}}{3\xi g^{2}}\right)
\;,\;\;\;  \nonumber \\
\varphi _{3}^{8} &=&-\sqrt{3}\varphi _{3}^{3},\;\overline{\varphi }_{3}^{8}=-%
\sqrt{3}\;\overline{\varphi }_{3}^{3}\;\;,  \label{r3}
\end{eqnarray}
with $M^{2}\;$being the renormalization scale and $\beta \overline{\beta }=1$%
. Expression $\left( \ref{veffsu3}\right) $ takes the same value for all
minima. As in the previous case of $SU(2)$, from the requirement that the
nontrivial vacuum configurations $\left( \ref{r1}\right) -\left( \ref{r3}%
\right) $ are solutions of the gap equation at weak coupling, for $\xi $ one
obtains the value 
\begin{equation}
\xi =\frac{2\beta _{0}}{3}=\frac{22}{3}\;.  \label{xisu3}
\end{equation}
It is worth underlining that the value obtained for $\xi $ is precisely the
same as in the case of $SU(2)$.

\section{Conclusion}

The existence of the off-diagonal ghost condensates $\left\langle
c\;c\right\rangle \;$and $\left\langle \overline{c}\;\overline{c}%
\right\rangle $ of dimension two in the MAG has been discussed. These
condensates rely on the nonlinearity of the MAG gauge fixing condition,
which requires the introduction of a quartic ghost self-interaction term,
needed for the renormalizability of the model. For positive values of the
gauge parameter $\xi $ the quartic self-interaction is attractive, favouring
the formation of ghost condensates, which show up as nontrivial solutions of
the gap equation for the effective potential.

A further important point is that the gauge parameter $\xi $ can be chosen
to ensure that the condensed vacuum configuration is a solution of the gap
equation at weak coupling, i.e. for small values of the gauge coupling
constant $g$. The whole framework is thus consistent with the perturbative
loop expansion. In particular, for $\xi $ one obtains the value $22/3$, for
both $SU(2)$ and $SU(3)$. Although we cannot extend the validity of this
mechanism to the strong coupling region, i.e. to energy scales below $%
\Lambda _{\mathrm{QCD}}$, it can be interpreted, to some extent, as a
possible evidence for the Abelian dominance. It is worth reminding indeed
that the off-diagonal condensate $\left\langle \overline{c}c\right\rangle $
is part of the more general condensate $\left( \frac{1}{2}\left\langle
A_{\mu }^{a}A^{\mu a}\right\rangle -\xi \left\langle \overline{c}%
^{a}c^{a}\right\rangle \right) ,$ which is expected to provide effective
masses for all off-diagonal fields \cite{ope,dd,dd1}.

In this work, the ghost condensation has been related to the dynamical
breaking of the ghost number symmetry, which is present in Yang-Mills theory
with arbitrary gauge group. It is useful to remind that the ghost number
generator $\delta _{FP}$ is part of $SL(2,R)$, which is present in the MAG
for any gauge group $SU(N)$ \cite{sl2r}$.$

Aspects concerning the Goldstone boson associated to this breaking as well
as the characterization of the condensate $\left( \frac{1}{2}\left\langle
A_{\mu }^{a}A^{\mu a}\right\rangle -\xi \left\langle \overline{c}%
^{a}c^{a}\right\rangle \right) \,$\thinspace are under investigation. We
remark that this massless excitation should be identified with a bound state
of ghosts. This follows by noting that, classically, the auxiliary fields $%
\varphi ^{i}$ and $\overline{\varphi }^{i}$ correspond to the ghost
composite operators $f^{iab}c^{a}c^{b}$ and $f^{iab}\overline{c}^{a}%
\overline{c}^{b}$. This massless excitation is expected to decouple from the
physical spectrum. In fact, in the case of $SU(2)$, a decoupling argument
based on the quartet mechanism \cite{kon} has been given for the Goldstone
boson related to the breaking of $SL(2,R)$ \cite{ms, ms1}.

Let us conclude with some general comments on the result so far obtained.
Certainly, many aspects of the ghost condensation remain to be analysed,
deserving a deeper understanding. Till now, the ghost condensates have been
investigated at one-loop order and in the weak coupling regime, where a
nontrivial vacuum seems to emerge. However, a complete analysis should
include a better understanding (at least qualitatively) of the strong
coupling. This would require to face genuine nonperturbative effects, such
as Gribov's ambiguities, which are also present in the Maximal Abelian Gauge 
\cite{gmag}. Here, the employment of the Schwinger-Dyson equations \cite{sd}
could provide more information about the role of the ghost condensates for
the infrared region of Yang-Mills theories.

Also, a two-loop analysis of the effective potential could improve our
understanding of the weak coupling regime and of the relationship $\left( 
\ref{v}\right) $ between the $\beta $-function and the gauge parameter $\xi $%
. The combined use of the local composite operators technique \cite{vkvv}
and of the algebraic renormalization \cite{book} proves to be particular
useful for this kind of analysis, as done in the case of the gluon-ghost
condensate $\left\langle \frac{1}{2}A^{2}-\xi \overline{c}c\right\rangle $
in the covariant nonlinear Curci-Ferrari gauge \cite{cfg}.

It is worth mentioning that by now evidences for the ghost condensation have
been reported in other gauges, namely in the Curci-Ferrari gauge \cite
{cfg1,cfg2,cfg3} and in the Landau gauge \cite{lg}. All these gauges,
including the Maximal Abelian Gauge, possess a global $SL(2,R)$ symmetry 
\cite{sl2r}, a feature which seems to be deeply related to the ghost
condensation.

Other important aspects to be further analysed are those related to the BCS\
versus Overhauser effect, \textit{i.e.} to establish which is the preferred
vacuum with the lowest energy. Also, the role of the BRST symmetry in the
presence of the ghost condensation needs to be clarified. We remark that
these aspects have been recently investigated in detail by \cite{rec}\ in
the case of the Curci-Ferrari and Landau gauge. Here, it turns out that, due
to the $SL(2,R)$ invariance of the effective potential, both the BCS\ and
the Overhauser vacua can be consistently chosen as vacuum state.
Furthermore, the resulting theory perturbed around the condensed vacuum is
found to be BRST\ invariant. A similar analysis is expected to apply in the
Maximal Abelian Gauge, leading essentially to the same conclusion.

\section*{Acknowledgements}

We are grateful to M.Schaden and D. Dudal for fruitful discussions. The
Conselho Nacional de Desenvolvimento Cient\'{\i }fico e Tecnol\'{o}gico
CNPq-Brazil, the Funda{\c{c}}{\~{a}}o de Amparo \`{a} Pesquisa do Estado do
Rio de Janeiro (Faperj) and the SR2-UERJ are acknowledged for the financial
support.{\ }

\end{document}